\documentclass[12pt,a4paper]{article}
\pdfoutput=1
\usepackage{jheppub}
\usepackage{tikz}
\usepackage{subcaption}
\DeclareMathOperator{\Tr}{Tr}

\newcommand{\ri}{\mathrm{i}}

\newcommand{\hf}{\frac{1}{2}}

\newcommand{\del}{\partial}

\newcommand{\bra}{\langle}
\newcommand{\ket}{\rangle}

\newcommand{\bt}{\beta}

\newcommand{\rt}[1]{\sqrt{#1}}

\newcommand{\cF}{\mathcal{F}}

\DeclareMathOperator{\arcsinh}{arcsinh}

\newcommand{\btot}{{\beta_\mathrm{tot}}}
\DeclareMathOperator{\Erf}{Erf}

\begin{document}

\begin{flushright}
	\hfill{OU-HET-1175}
\end{flushright}
\title{\mathversion{bold}Late time behavior of $n$-point spectral form factors  in Airy and JT gravities}

 \author[a]{Takanori Anegawa,}
 \author[a]{Norihiro Iizuka,}
\author[b]{Kazumi Okuyama}
\author[c]{and Kazuhiro Sakai}

\affiliation[a]{\it Department of Physics, Osaka University, 
Toyonaka, Osaka 560-0043, JAPAN} 
\affiliation[b]{Department of Physics, Shinshu University,\\
3-1-1 Asahi, Matsumoto 390-8621, JAPAN}
\affiliation[c]{Institute of Physics, Meiji Gakuin University,\\
1518 Kamikurata-cho, Totsuka-ku, Yokohama 244-8539, JAPAN}

\emailAdd{takanegawa@gmail.com}
\emailAdd{iizuka@phys.sci.osaka-u.ac.jp}
\emailAdd{kazumi@azusa.shinshu-u.ac.jp}
\emailAdd{kzhrsakai@gmail.com}

\abstract{
We study the late time behavior of $n$-point spectral form factors (SFFs) in two-dimensional Witten-Kontsevich topological gravity, which includes both Airy and JT gravities as special cases. This is conducted in the small $\hbar$ expansion, where $\hbar \sim e^{- {1}/{G_N}}$ is the genus counting parameter and nonperturbative in Newton's constant $G_N$. For one-point SFF,  we study its absolute square at two different late times. We show that it decays by power law at $t \sim \hbar^{-2/3}$ while it decays exponentially at $t \sim \hbar^{-1}$ due to the higher order corrections in $\hbar$. We also study general $n (\ge 2)$-point SFFs at $t \sim \hbar^{-1}$ in the leading order of the $\hbar$ expansion. We find that they are characterized by a single function, which is essentially the connected two-point SFF and is determined by the classical eigenvalue density $\rho_0(E)$ of the dual matrix integral. These studies suggest that qualitative behaviors of $n$-point SFFs are similar in both Airy and JT gravities, where our analysis in the former case is based on exact results. 
}

\maketitle

\section{Introduction}

One of the goals in quantum gravity is to understand the microscopic
structures of black hole spacetime. Given the AdS/CFT correspondence
\cite{Maldacena:1997re}, in principle, one can understand the quantum
spectrum of spacetime by directly solving the large $N$ CFT.
However, the large $N$ CFTs in general are notoriously difficult to solve, and therefore it is in practice almost impossible to determine the microscopic states.   

In recent years, as a simple toy model for AdS/CFT, the low-energy
correspondence between a specific two-dimensional dilaton gravity,
Jackiw-Teitelboim (JT) gravity \cite{Teitelboim:1983ux, Jackiw:1984je},
and the one-dimensional Majorana fermion model, Sachdev-Ye-Kitaev (SYK)
model \cite{Sachdev:1992fk, Kitaevtalk} has received a great deal of attention. 
Particularly noteworthy is that the low-energy reparametrization modes
of both JT gravity and the SYK model  
 are described by the same one-dimensional Schwarzian action \cite{Maldacena:2016hyu, Maldacena:2016upp}, which shows the maximal chaos bound \cite{Maldacena:2015waa} expected from a black hole \cite{Sekino:2008he}. 
 
At the same time, there has been significant progress in understanding the effects of non-perturbative effects in Newton's constant $G_N$, as exemplified by the finding of the island formula \cite{Penington:2019npb, Almheiri:2019psf, Almheiri:2019hni} and replica wormholes \cite{Penington:2019kki, Almheiri:2019qdq}. The key point is that 
there are non-perturbative new saddle points in the gravitational path integral, 
which are irrelevant in the early time, but in the late time these new saddles become 
relevant and play a very important role for the entanglement entropy of the Hawking radiations showing the Page curve \cite{Page:1993df, Page:1993wv}. Here early and late is in comparison with the Page time.

The reason why the late time is important must be clear: in the late
time, the detailed energy eigenvalue distribution of microscopic
theories plays an important role. Especially whether energy eigenvalues have
correlations or not is essential: typically for chaotic systems, the energy
eigenvalues repel \cite{Bohigas:1983er}, and are described by the ``sine-kernel'' 
\cite{Gaudin:1961,Dyson:1962es}. In fact,
this property can be seen directly by the study of the two-point
spectral form factor (SFF)  in the SYK model \cite{Cotler:2016fpe},
which is defined as the analytically continued
correlator of partition functions 
\begin{align}
\langle Z(\beta+\ri t)Z(\beta-\ri t) \rangle,
\end{align}
where $\langle \cdot \rangle$ denotes the disorder average for 
the SYK model.  
In the SYK model, this two-point SFF shows the decay (called slope) in
the early time, but in the late time it  shows the time-linear growth called
{\it ramp} and 
the constant behavior called {\it plateau}. 
 The ramp and plateau are seen in   
the random matrix theory 
 (RMT) as well
and therefore 
are regarded as typical behaviors of chaotic systems.

Recently, in a beautiful paper \cite{Saad:2019lba} it has been shown that 
given the leading classical eigenvalue density $\rho_0(E) \sim \sinh \sqrt{E}$ \cite{Stanford:2017thb},  
there is an exact correspondence between JT gravity and RMT in the perturbation expansion; one in terms of $e^{-S} = e^{-1/G_N}$, where $S$ is the entropy, and the other in terms of $1/L$, where $L$ is the rank of RMT. 
This is based on the equivalence of the recursion relation between the
Mirzakhani's one for Weil-Petersson volumes
\cite{Mirzakhani:2006fta} and the topological recursion of RMT by Eynard and Orantin \cite{Eynard:2007fi}. 
This reveals that there are intertwined relationships  
between SYK models, RMT, and JT gravity. 
Furthermore, it has been pointed out in \cite{Okuyama:2019xbv} that JT gravity is a special case of 
more general two-dimensional Witten-Kontsevich (WK) topological
gravity. WK topological gravity contains 
infinitely many couplings $t_k$, and by tuning all of $t_k$ to specific values,  it reproduces the classical eigenvalue density $\rho_0(E) \sim \sinh \sqrt{E}$ of JT gravity.

Historically two-dimensional topological gravity, where observables are intersection numbers in the moduli space of Riemann surfaces, has been extensively studied from 80's \cite{Brezin:1977sv, Brezin:1990rb, Douglas:1989ve, Gross:1989vs}.   
It was conjectured by Witten \cite{Witten:1990hr} and proved by
Kontsevich \cite{Kontsevich:1992ti} that this topological gravity is
equivalent to the double-scaled one matrix model which counts
triangulations of surfaces and similarly contains infinitely many couplings associated to the matrix potential $V$.  
Based on this equivalence, systematic studies on multi-boundary partition
functions in two-dimensional topological gravity, which encompasses JT
gravity, have been 
done in \cite{Okuyama:2020ncd}. The key points are that the
generating function for the intersection numbers of topological gravity
is known to obey the Korteweg-De Vries (KdV) equation and can be obtained by genus expansion. 
Once the multi-boundary partition functions in two-dimensional
topological gravity are obtained,
by the analytic continuation $\beta_i\to\beta_i+\ri t_i$ 
one can also obtain generic $n$-point SFFs in topological gravity. The case of $n=2$ especially corresponds 
 to the conventional SFF studied before \cite{Cotler:2016fpe}. 
It is a natural question whether the full behavior of the SFF seen in SYK and RMT, especially 
the ramp and plateau can be reproduced from JT gravity or more generically topological WK gravity. 
Recently it has been shown in JT gravity that  by taking an appropriate rescaling for the late time, one
can conduct the summation of the Weil-Petersson volumes 
for arbitrary genus $g$ and see not only the ramp, but also its transition to
plateau as well \cite{Blommaert:2022lbh, Saad:2022kfe, Okuyama:2023pio}.

Given the success of $n=2$ SFF,  in this paper we study more generic
$n$-point SFFs in late time
for generic WK topological gravity which includes JT gravity. 
In RMT, $n$-point generalizations of 2-point SFF in RMT for $n$ even 
 were studied in \cite{Cotler:2017jue, Liu:2018hlr}. 
In this paper, we study general
$n$-point SFFs in topological gravity, {\it i.e.} after the double scaling limit of RMT. 
Furthermore, in our studies
$n$ can be odd, in fact we study the $n=1$ case as well. 
By examining the $n$-point SFF with analytically continued time in the late time limit, 
we attempt to understand its universal behavior in topological gravity. 

Especially we elucidate 
how to take the late time limit
of the one-point SFF in generic two-dimensional WK topological gravity
in the small $\hbar$ expansion, where 
\begin{align}
\hbar \sim e^{-S} \sim e^{- \frac{1}{G_N}}
\end{align} 
is the genus-counting parameter. Note that $\hbar$ corrections are non-perturbative effects in $G_N$. 
We analyze two different late times, 
one in $t \sim \hbar^{-2/3}$ and
the other in $t \sim \hbar^{-1}$ in the $\hbar \to 0$ limit.
We will see that at relatively late time $t \sim \hbar^{-2/3}$ and very
late time $t \sim \hbar^{-1}$, the behavior of
the one-point SFF is different. 

We also present a systematic analysis of $n$-point SFF at $t \sim \hbar^{-1}$.
We find that they are characterized by a single
function, which is essentially the connected two-point SFF and
is determined by the classical eigenvalue density
$\rho_0(E)$ of the dual matrix integral. 
Furthermore, not only generic topological WK gravity, 
we also consider 
Airy and JT gravities as concrete examples,
both of which are obtained by tuning the
infinitely many couplings $t_k$ in WK topological gravity. We see that both Airy and JT
gravities behave in a qualitatively very similar way, 
 which we can see from generic topological WK gravity. From these, we
 conjecture that the qualitative behavior of the $n$-point SFF for generic
 topological gravity including JT gravity 
is very similar to that of Airy gravity. 
In other words, if one ask how special Airy or JT gravity is,  
our temporal answer is that most probably 
Airy and JT gravities are quite typical in generic topological WK gravity parameter range.

The organization of this paper is as follows; in \S\ref{sec:airy} we
study the SFF for Airy gravity where the exact full $n$-point 
function is known. In \S\ref{sec:1pt} 
we study the generic behavior of the one-point SFF
in general WK topological gravity, 
including JT gravity as one of the concrete examples. 
In \S\ref{sec:n-point}
we study $n$-point SFF in general WK topological gravity
at very late times $t\sim\hbar^{-1}$. 
We also study the results of JT gravity $n$-point SFF as
a concrete example.
We end with conclusions and discussion at \S\ref{sec:discussion}.

\section{\mathversion{bold} $n$-point spectral form factor in Airy gravity
}\label{sec:airy}
\subsection{Airy gravity overview}\label{sec:airyoverview}

Exact general $n$-point partition functions are difficult to obtain in
generic WK topological gravity. Therefore 
we are obliged to employ some sort of expansions 
 \cite{Okuyama:2020ncd}. 
However, there is an exception; which is Airy gravity.   
Airy gravity corresponds to the one matrix model which is given by Gaussian potential, 
therefore it is solvable. In fact, general $n$-point function is given exactly  
in the integral form, which was worked out first by Okounkov \cite{Okounkov:2001usa}.\footnote{See also \cite{Ginsparg:1993is,Maldacena:2004sn} as well as the appendix A of \cite{Okuyama:2019xbv}.}

The Airy gravity is obtained by zooming in on the edge of the Wigner
semi-circle: its classical eigenvalue density is given by\footnote{Our
convention for the eigenvalue density is different by $2 \pi \hbar$ than \cite{Okuyama:2019xbv}.}
\begin{align}
{\rho_0(E) = 2\sqrt{E}.}
\label{eq:rho0-airy}
\end{align}
The double-scaled wave function $\psi(E)$, which is a Baker-Akhiezer function of the KdV hierarchy, 
obeys the Schr\"{o}dinger equation: 
\begin{align}
(\hbar^2\del_x^2+x+E)\psi (E)=0 \,.
\label{eq:AiryODE}
\end{align} 
Here $x$ and $E$ are continuous parameters obtained by the 
double-scaled limit of polynomial index $n$ and matrix eigenvalue $\lambda$ respectively.  
We also introduce the notation
\begin{align}
\psi(E) \equiv \bra x|E\ket,\qquad Q:=\hbar^2\partial_x^2+x,
\end{align}
where $|x\ket$ is the coordinate eigenstate
and $|E\ket$ is the energy eigenstate satisfying
$Q|E\ket = -E|E\ket$. 
The Baker-Akhiezer function in the present case is written in terms of 
the Airy function
\begin{equation}
\begin{aligned}
\psi(E) =\hbar^{-\frac{2}{3}}\text{Ai}\bigl[-\hbar^{-\frac{2}{3}}(E+x)\bigr].
\end{aligned} 
\label{eq:BA-airy}
\end{equation}

Given the Baker-Akhiezer function $\psi(E) \equiv \bra x|E\ket$, one can obtain the full eigenvalue density  as 
\begin{equation}
\begin{aligned}
{\frac{\rho_{\text{Airy}}(E)}{2\pi\hbar}}
=\int_{-\infty}^0 dx  \bra x|E\ket^2 = \hbar^{-\frac{2}{3}} \left[ \text{Ai}'\bigl(\zeta \bigr)^2 -  \zeta \text{Ai}\bigl(\zeta \bigr)^2  \right] \,, 
\end{aligned} 
\label{eq:rho-airy-int}
\end{equation}
where $\zeta = -\hbar^{-\frac{2}{3}}E$ and the Airy function satisfies $\text{Ai}''\bigl(\zeta \bigr) = \zeta \text{Ai}\bigl(\zeta \bigr)$. This defines a non-perturbative completion of the classical
eigenvalue density \eqref{eq:rho0-airy}. In fact, using the asymptotic formula for the Airy function 
\begin{align}
  \text{Ai}\bigl(\zeta \bigr)  \sim \frac{       \cos \left( \frac{\pi}{4} - \frac{2 |\zeta|^{\frac{3}{2}}} {3} \right)}{\sqrt{\pi} |\zeta|^{\frac{1}{4}}} \quad \mbox{(at $ \zeta \to -\infty$)} \,,
\end{align}
one can see that in the $\hbar \to 0 $ limit, $\rho_{\text{Airy}}(E)$
given by  eq.~\eqref{eq:rho-airy-int} reduces to the classical $\rho_0(E) $ given by eq.~\eqref{eq:rho0-airy}. 

The one-point function of the macroscopic loop operator is given by 
\begin{equation}
\begin{aligned}
 \bra Z(\bt)\ket&=
\int_{-\infty}^0 dx \bra x|e^{\bt Q}|x\ket=
\int_{-\infty}^\infty dE e^{-\bt E}
{\frac{\rho_{\text{Airy}}(E)}{2\pi\hbar}}
 =\frac{e^{\frac{\hbar^2\bt^3}{12}}}{2\rt{\pi}\hbar\bt^{3/2}}.
\end{aligned} 
\label{eq:airy-one}
\end{equation}
In the planar limit, this one-point function behaves as 
\begin{align}
 \bra Z(\bt)\ket&=  \bra Z(\bt)\ket_0 +  O(\hbar^2),\\
 \bra Z(\bt)\ket_0 &\equiv \frac{1}{2\rt{\pi}\hbar\bt^{3/2}}.
 \label{planaronepoint}
\end{align}
Here the subscript $0$ represents the genus zero planar limit.

\subsection{One point function: $n=1$ SFF}
We start our analysis  from the one-point function of analytically continued partition function in Airy gravity. For that purpose, in eq.~\eqref{eq:airy-one}, 
we analytically continue $\beta \to \beta + \ri   t$ and obtain  
\begin{align}
 \bra Z(\bt + \ri t)\ket&= \frac{e^{\frac{\hbar^2 \left( \bt + \ri t \right)^3}{12}}}{2\rt{\pi}\hbar \left(\bt + \ri t \right)^{3/2}} \,,
 \label{fullanalyticconin}
 \\
 \bra Z(\bt + \ri t )\ket_0 &= \frac{1}{2\rt{\pi}\hbar \left(\bt + \ri t \right)^{3/2}}  \,.
 \label{planaranalyticcontin}
\end{align}

Furthermore, by normalizing it at the value of $t=0$,  the exact one-point function becomes 
\begin{align}
\frac{\langle Z(\beta+\ri t) \rangle}{\langle Z(\beta) \rangle} 
= \frac{e^{\frac{\hbar^2 (\beta+\ri t)^3}{12}}}{e^{\frac{\hbar^2 \beta^3}{12}}}  \cdot \frac{\beta^{3/2}}{(\beta+\ri t)^{3/2}}.
\end{align}
On the other hand, in the planar limit, where we keep only the leading
order part in the $\hbar \to 0$ limit, we have   
\begin{align}
\frac{\langle Z(\beta+\ri t) \rangle}{\langle Z(\beta) \rangle} = 
\frac{\beta^{3/2}}{(\beta+\ri t)^{3/2}}+O(\hbar^2 (\beta + \ri t)^3 ) = \frac{\langle Z(\beta+\ri t) \rangle_0}{\langle Z(\beta) \rangle_0}+O(\hbar^2 (\beta + \ri t)^3 ),
\end{align}
where $\langle Z(\beta) \rangle_0$ is given by
eq.~\eqref{planaronepoint}
and 
we implicitly assumed that  
\begin{align}
\hbar^2 (\beta + \ri t)^3  \ll 1  \,.
\label{gssmall}
\end{align}

\begin{figure}[htbp]
 \centering
 \includegraphics[keepaspectratio, scale=1.15]{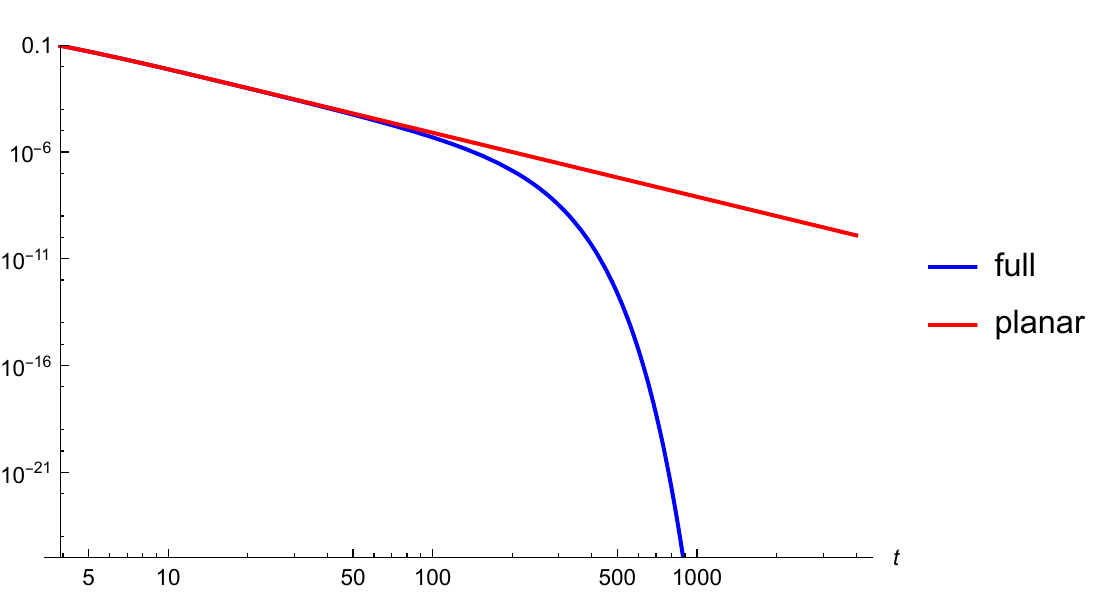}
 \caption{Log-log plot of the  full and planar disconnected two
 point SFFs given by eq.~\eqref{fullairydisc} and \eqref{zerodisc}
 respectively. 
Here we set $\beta=2, \hbar=0.01/\sqrt{2}$.
The effect of higher order corrections comes into view
at $t=O(t_{\rm higher \, genus}) = O\left((\hbar\sqrt{\beta})^{-1}\right)= O(100)$.}
\label{airydisc}
\end{figure}

To get rid of the phase factor, let us compare the planar contribution $
\frac{ \langle Z(\beta+\ri t)  \rangle_0 }{ \langle Z(\beta) \rangle_0
}$ with the full contribution $\frac{\langle Z(\beta+\ri t)
\rangle}{\langle Z(\beta) \rangle} $ by taking their absolute 
square. 
This yields the  disconnected two-point SFF 
\begin{align}
\frac{ \langle Z(\beta+\ri t) \rangle \langle Z(\beta-\ri t) \rangle }{\langle Z(\beta) \rangle^2}
=\frac{\beta^3}{(\beta^2+t^2)^{3/2}} e^{\frac{- \hbar^2 \beta }{2} t^2}.
\label{fullairydisc}
\end{align}
Since the planar part is 
\begin{align}
\frac{\langle Z(\beta+\ri t) \rangle_0 \langle Z(\beta-\ri t) \rangle_0 }{\langle Z(\beta) \rangle_0^2}=\frac{\beta^3}{(\beta^2+t^2)^{3/2}},
\label{zerodisc}
\end{align}
the difference between the planar and the full contribution is
whether the exponential factor can be neglected or not,  
\begin{align}
\frac{\langle Z(\beta+\ri t) \rangle \langle Z(\beta-\ri t) \rangle }{\langle Z(\beta) \rangle^2} : \frac{\langle Z(\beta+\ri t) \rangle_0 \langle Z(\beta-\ri t) \rangle_0 }{\langle Z(\beta) \rangle_0^2} = e^{\frac{-\hbar^2 \beta t^2}{2}} : 1.
\end{align}
 Thus, higher order corrections in $\hbar$ become important at the time-scale of 
\begin{align}
\hbar^2 \beta t^2 = O(1) &\Leftrightarrow
 t  = O(t_{\rm higher \, genus})\, , \\ 
\mbox{where}  \quad  &t_{\rm higher \, genus} \equiv \frac{1}{\hbar \sqrt{\beta}} \,.
\label{defthiger}
\end{align}
At $t= O(t_{\rm higher \, genus})$, higher genus corrections become
important (see Figure~\ref{airydisc}).
This time-scale $t_{\rm higher \, genus}$ is called $t_{\rm plateau}$ in \cite{Okuyama:2019xbv}, the time-scale where ramp changes into plateau. 

\subsection{Connected two-point spectral form factor}

Next we consider the  two-point SFF for Airy gravity, which is obtained by analytically continuing 
the connected correlator of two macroscopic loops
\begin{equation}
\begin{aligned}
 \bra Z(\bt_1)Z(\bt_2)\ket_{\text{conn}}&=\Tr \big(e^{\bt_1Q}(1-\Pi)e^{\bt_2Q}\Pi\big) \,, 
\end{aligned} 
\end{equation}
where $\Pi$ is the projector 
\begin{equation}
\begin{aligned}
 \Pi=\int_{-\infty}^0 dx|x\ket\bra x|.
\end{aligned} 
\end{equation}

\begin{figure}[t]
 \centering
 \includegraphics[keepaspectratio, scale=1.2]{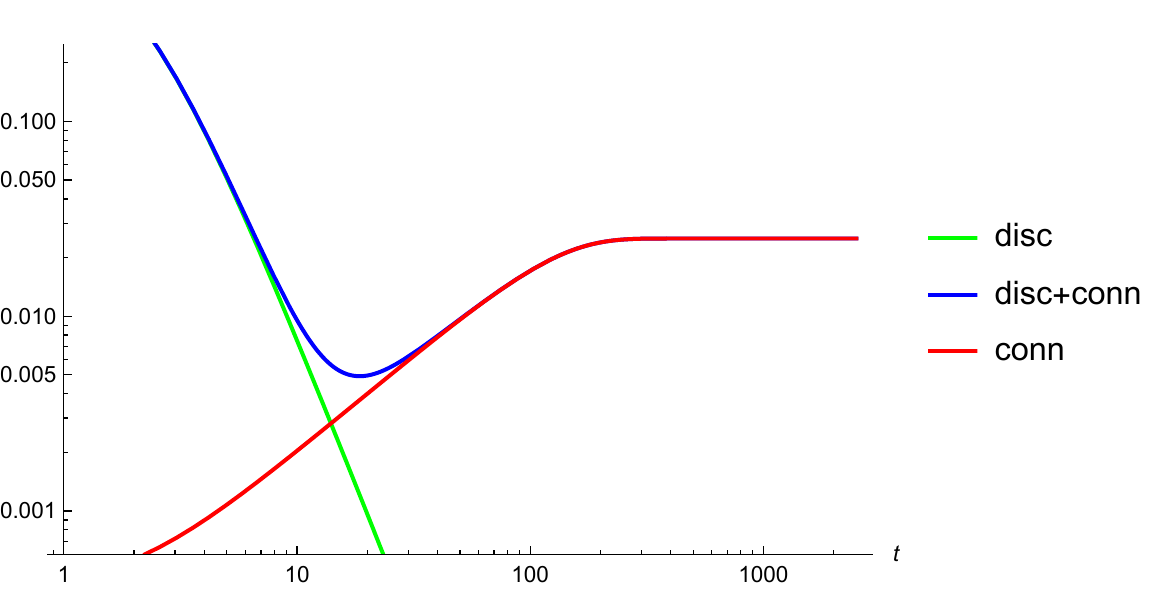}
 \caption{Log log plot of the  full SFF (blue) given as the  Sum
 of eq.~\eqref{fullairydisc} and eq.~\eqref{fullairyconn}, compared with
 the disconnected SFF (green) given by eq.~\eqref{fullairydisc} and
 the connected SFF (red) given by eq.~\eqref{fullairyconn}  as a
 function of time $t$. In this figure, we take $\beta=2,
 \hbar=0.01/\sqrt{2}$, so that the dip time is 
$t_{\rm dip}\sim \beta^{1/4}/\sqrt{\hbar}\sim 14$.
One can see the slope ($t \lesssim 14$), dip ($t\sim 14$),
ramp ($14 \lesssim t \lesssim 100$) and plateau ($t \gtrsim 100$)
in Airy gravity.}
\label{airyfull}
\end{figure}

\begin{figure}[h]
 \centering
 \includegraphics[keepaspectratio, scale=1.2]{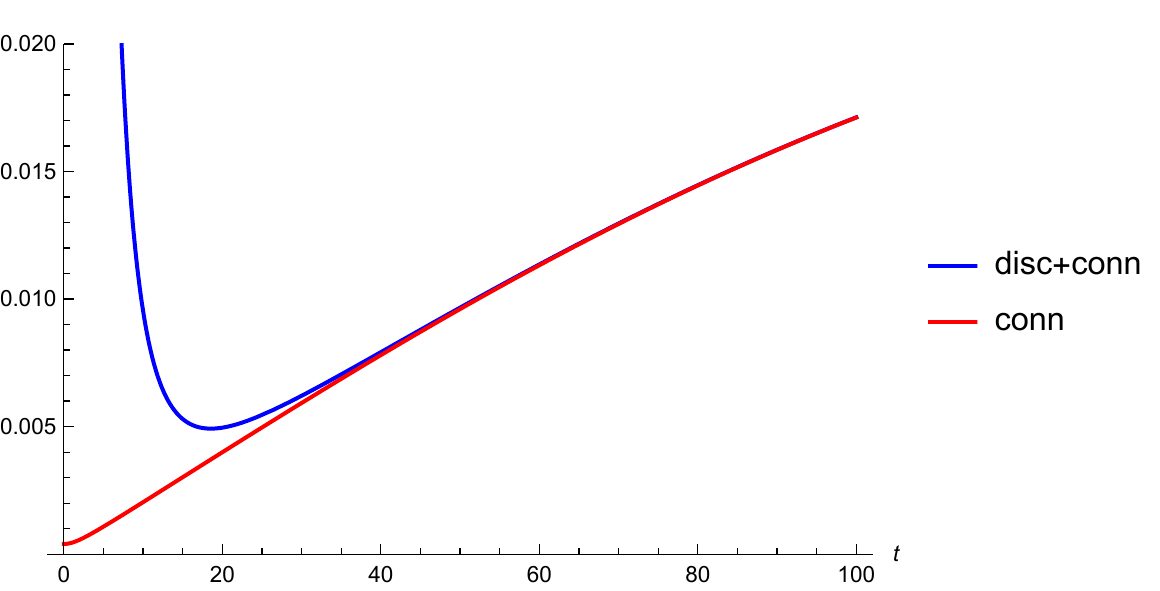}
 \caption{Plot of the same functions as in Figure \ref{airyfull}, with more 
focus on the ramp region $ t \lesssim 100$.
One observes the $t$-linear growth, which is 
due to the eigenvalue repulsion.}
\label{airyfull2}
\end{figure}

The  full connected two-point function is written in terms of the  error function as 
\begin{align}
\langle Z(\beta_1)Z(\beta_2) \rangle_{\rm conn}
=\frac{e^{\frac{\hbar^2 (\beta_1+\beta_2)^3}{12}}}{2 \sqrt{ \pi}\hbar (\beta_1+\beta_2)^{3/2}}{\rm Erf}\left(\frac{\hbar}{2}\sqrt{\beta_1\beta_2 (\beta_1+\beta_2)}\right) \,.
\label{twopointconnbeta12}
\end{align}
By analytically continuing the arguments as 
\begin{align}
\label{beta1beta2rela}
\beta_1 \equiv \beta + \ri t \,, \quad \beta_2 \equiv \beta - \ri t
\end{align}
and normalizing by the one point function  
 at $t=0$, the full two-point function is
\begin{align}
\frac{\langle Z(\beta+\ri t)Z(\beta-\ri t) \rangle_{\rm conn}}{\langle Z(\beta) \rangle^2}&
= 
\left(\frac{e^{\frac{\hbar^2 \beta^3}{12}}}{2 \sqrt{ \pi} \hbar \beta^{3/2}}\right)^{-2}
\cdot \frac{e^{\frac{\hbar^2 ( 2 \beta)^3}{12}}}{2 \sqrt{ \pi} \hbar ( 2 \beta)^{3/2}}{\rm Erf}\left(\frac{\hbar}{2}\sqrt{2 \beta (\beta^2 + t^2) }\right) \nonumber \\
&=\sqrt{ \frac{ \pi }{2} } \hbar \beta^{3/2} e^{\frac{\hbar^2 \beta^3}{2}} {\rm Erf}\left(\frac{\hbar}{2}\sqrt{2 \beta(\beta^2+t^2)}\right)  \,.
\label{fullairyconn}
\end{align}
Note that this does not go to zero at $t=0$,
since we have normalized the connected function by the disconnected one. 
In Figures~\ref{airyfull} and \ref{airyfull2}, we plot the connected SFF given in eq.~\eqref{fullairyconn}  and connected plus disconnected SFF for $\beta = 2$, $\hbar =0.01/\sqrt{2}$. 

\begin{figure}[tbp]
 \centering
 \includegraphics[keepaspectratio, scale=1.2]{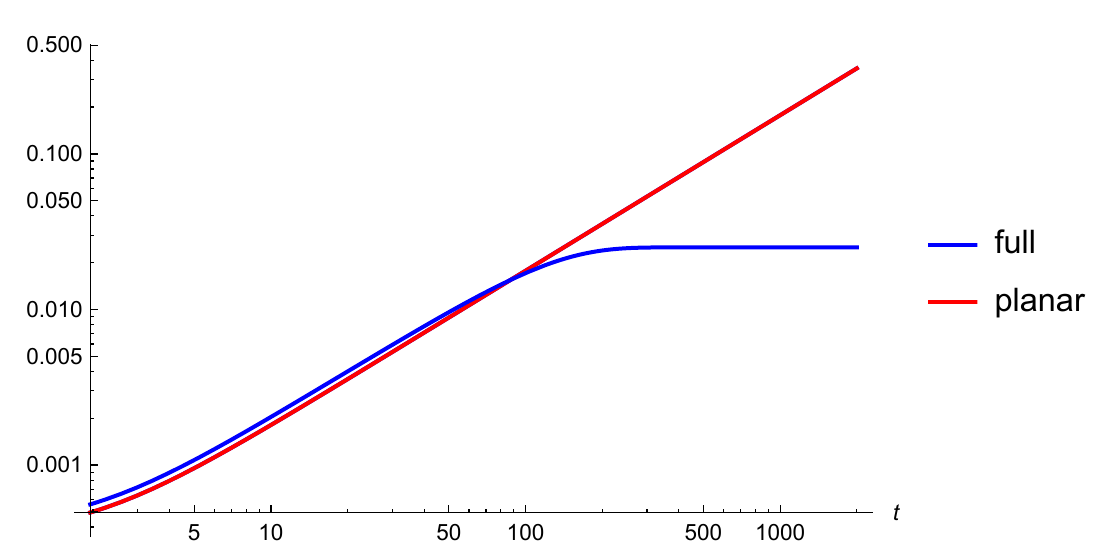}
 \caption{Comparison between the full two-point function given by eq.~\eqref{fullairyconn} and its planar limit given by eq.~\eqref{connzero}. 
The deviation starts at $t_{\rm higher \, genus} $ due to higher order 
corrections, which change
 ramp into plateau. 
We take $\beta=2, \hbar=0.01/\sqrt{2}$, so that 
$t_{\rm higher \, genus}=(\hbar \sqrt{\beta})^{-1} = 100$.}
\label{airyconnfull}
\end{figure}

In the planar limit $\hbar \to 0$, we have 
\begin{align}
e^{\frac{\hbar^2 \beta^3}{2}} =  1 + O(\hbar) \,, \quad {\rm Erf}\left(\frac{\hbar}{2}\sqrt{2 \beta(\beta^2+t^2)}\right) =  \frac{\hbar}{2}\sqrt{2 \beta(\beta^2+t^2)} + O(\hbar^2) \,. 
\end{align}
Therefore the planar contribution to the above connected two-point function is 
\begin{align}
\frac{\langle Z(\beta+\ri t)Z(\beta-\ri t) \rangle_{\rm conn}}{\langle Z(\beta) \rangle^2}
=  \frac{\sqrt{\pi}}{2} \hbar^2 \beta^2\sqrt{\beta^2+t^2} + O(\hbar^3) \,, 
\end{align}
and at $t \gg \beta$, its planar contribution is proportional to time $t$, 
\begin{align}
\frac{\langle Z(\beta+\ri t)Z(\beta-\ri t) \rangle_{\rm conn \, 0}}{\langle Z(\beta) \rangle_0^2} \equiv
\frac{\sqrt{\pi}}{2} \hbar^2 \beta^2\sqrt{\beta^2+t^2}  \propto \hbar \beta^2 t \,.
\label{connzero}
\end{align}
At late times $t  \gg \beta$, 
the argument of the error function becomes of order one  
\begin{align}
\hbar \sqrt{\beta t^2}= O( 1) \Leftrightarrow t = O\left( t_{\rm higher \, genus} \right) .
\end{align}
Therefore higher order corrections become important
at $t_{\rm higher \, genus}$, given in eq.~\eqref{defthiger}
(see Figure~\ref{airyconnfull}). 
Note that this typical time-scale $t_{\rm higher \, genus}$ where
higher genus contribution becomes important is the same
for both one-point and connected two-point functions.

At $t= O( t_{\rm higher \, genus})$, time-dependence changes;  
for one-point function, the way of decay
changes from power law decay to exponential decay due to higher genus
corrections and for two-point function,
higher genus corrections change 
the ramp into plateau.  Note also that the signature of ramp and plateau is common for both Airy and JT gravities \cite{Cotler:2016fpe}.

\subsection{Dip time $t_{\rm dip}$ vs  $ t_{\rm higher \, genus}$}

Before we proceed to the three point function, we comment on the dip time $t_{\rm dip}$. 
Dip time is calculated as the time where the magnitude of the connected
two-point SFF becomes of the same order as  
 the disconnected contribution. Generically $t_{\rm dip} \ll t_{\rm plateau}$ as we will see. 

Since $t_{\rm plateau}$ is the timescale where higher genus corrections become important, if  $t_{\rm dip} \ll t_{\rm plateau}$, at the dip time, all exponential factor can be approximated as one, in other words,  the lowest genus contribution dominates due to the definition of $t_{\rm plateau}$.  
The disconnected part of the two point function in the planar limit
is given by eq.~\eqref{zerodisc}. 
Equating this with eq.~\eqref{connzero}, we have 
\begin{align}
\frac{\beta^3}{(\beta^2+t_{\rm dip}^2)^{3/2}}
\sim
\hbar^2 \beta^2\sqrt{\beta^2+t_{\rm dip}^2}.
\end{align}
From this,  we obtain the dip time  
\begin{align}
 t_{\rm dip} \equiv \frac{\beta^{1/4}}{\sqrt{\hbar}}  \,.
\end{align}

If the temperature is
of order one, {\it i.e.,} $\beta = O(1)$, 
in the limit of $\hbar \to 0 $, the dip time and higher genus time are  
\begin{align}
t_{\rm dip} &\sim \hbar^{-1/2} \,, \quad \\
t_{\rm higher \, genus} &\sim \hbar^{-1} \,,
\label{highergeneshbar-1}
\end{align}
and there is a large hierarchy between $t_{\rm dip}$ and $t_{\rm higher \, genus}$, 
\begin{align}
t_{\rm dip}\ll t_{\rm higher \, genus} \,.
\end{align}
As we will analyze in detail later, the timescale $t \sim \hbar^{-1}$ is
where higher genus effects are important and simultaneously one can
analyze the SFF using what is called the $\tau$-scaling limit.

These results in Airy gravity show that 
the behavior of SFF in Airy gravity is very similar to that of JT gravity 
\cite{Okuyama:2020ncd, Blommaert:2022lbh, Saad:2022kfe, Okuyama:2023pio}. 
The SFF in Airy gravity shows ramp and plateau just as in JT gravity and the hierarchy between ${t_{\rm dip}}$ and ${t_{\rm higher \, genus}}$ is  
\begin{align}
\frac{t_{\rm dip}}{t_{\rm higher \, genus}} \sim {\hbar^{1/2}} 
\end{align}
with $\hbar \sim e^{- S}$, which is the same as seen in both JT gravity and RMT in  \cite{sff1986prl, Cotler:2016fpe}.

\subsection{Three-point spectral form factor}

Even though the general $n$-point function of
$Z(\bt)$'s in Airy gravity was obtained by Okounkov in the form
of an $n$-dimensional integral \cite{Okounkov:2001usa}, 
this integral is still very complicated and the closed form expression
of the $n$-point function for $n\geq4$ is not known in the literature,
as far as we are aware of. However for $n=3$, the closed form
expression was obtained in
\cite{Okuyama:2021cub} as
\begin{equation}
\begin{aligned}
 &\frac{\bra Z(\bt_1)Z(\bt_2)Z(\bt_3)\ket_{\text{conn}}}{\bra Z(\bt_1+\bt_2+\bt_3)\ket}\\
& =1-4T\left(\frac{ \hbar}{\sqrt{2}}\rt{\bt_1(\bt_2+\bt_3)(\bt_1+\bt_2+\bt_3)},
\rt{\frac{\bt_2\bt_3}{\bt_1(\bt_1+\bt_2+\bt_3)}}\right)\\
& \qquad -4T\left(\frac{ \hbar}{\sqrt{2}}\rt{\bt_2(\bt_3+\bt_1)(\bt_1+\bt_2+\bt_3)},
\rt{\frac{\bt_3\bt_1}{\bt_2(\bt_1+\bt_2+\bt_3)}}\right)\\
& \qquad -4T\left(\frac{ \hbar}{\sqrt{2}}\rt{\bt_3(\bt_1+\bt_2)(\bt_1+\bt_2+\bt_3)},
\rt{\frac{\bt_1\bt_2}{\bt_3(\bt_1+\bt_2+\bt_3)}}\right),
\end{aligned} 
\label{eq:Airy-3pt}
\end{equation}
where $T(z,a)$ denotes the Owen's $T$-function
\begin{equation}
\begin{aligned}
 T(z,a)=\frac{1}{2\pi}\int_0^adx\frac{e^{-\hf z^2(1+x^2)}}{1+x^2} \,.
\end{aligned} 
\end{equation}

As before, let us consider the analytic continuation of $\bt_i$
to a complex value,
\begin{equation}
\begin{aligned}
 \bt_i\to \bt_i+\ri t_i,\quad (i=1,2,3).
\end{aligned} 
\end{equation}
We can focus on the late time behavior of the three-point function
\eqref{eq:Airy-3pt} around the time scale of plateau
by taking the $\tau$-scaling limit
\begin{equation}
\begin{aligned}
 t_i\to \infty, \quad \hbar\to0,\quad \tau_i=t_i\hbar=\text{finite}.
\end{aligned} 
\end{equation}
It turns out that  
the late time behavior is quite different for $\sum_i \tau_i=0$ and $\sum_i \tau_i\ne0$.

First, let us consider the case $\sum_i \tau_i=0$.
Without loss of generality, we can assume
\begin{equation}
\begin{aligned}
 \tau_1,\tau_2>0,\quad \tau_3=-(\tau_1+\tau_2)<0.
\end{aligned} 
\end{equation}
Using the property of the Owen's $T$-function, one can show that
in the $\tau$-scaling limit the three-point function becomes
\begin{equation}
\begin{aligned}
 \left\bra\prod_{i=1}^3Z(\bt_i+\ri\tau_i\hbar^{-1})\right\ket_{\text{conn}}
= 
\langle Z(\bt_{\text{tot}}) \rangle_0
\sum_{i=1}^3\text{Erf}\left(\hf\tau_i\rt{\bt_{\text{tot}}}\right),
\end{aligned} 
\label{eq:Airy-3pt-late}
\end{equation}
where $\bt_{\text{tot}}=\sum_i\bt_i$ and 
$\langle Z(\beta) \rangle_0$ is the planar limit of one-point function given by eq.~\eqref{planaronepoint}.

\begin{figure}[t]
\centering
\includegraphics
[width=1\linewidth]{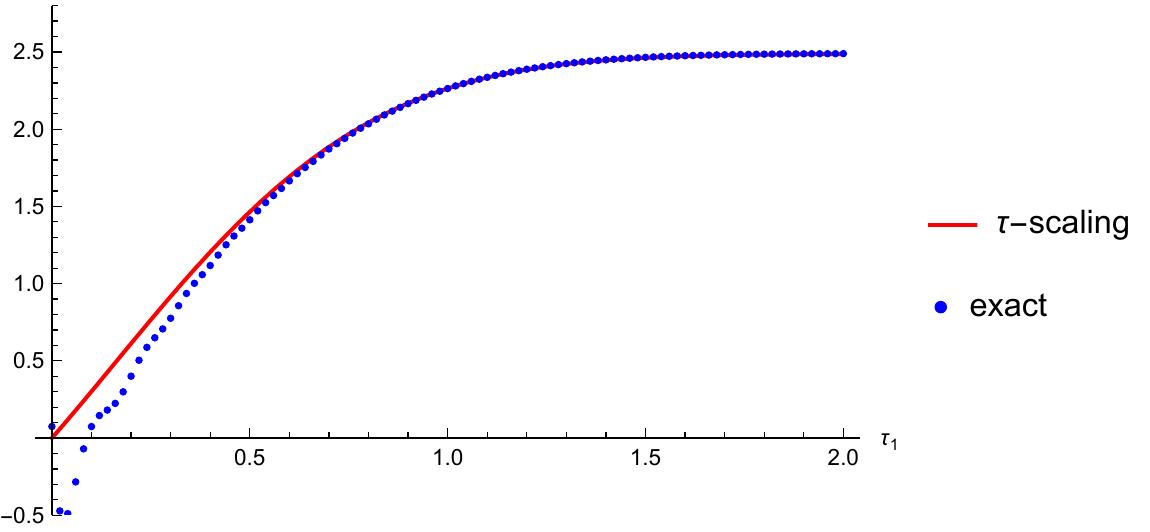}
\caption{Plot of the three-point function for $\sum_i\tau_i=0$
as a function of $\tau_1$ with fixed $\tau_2$.
In this figure we set $\hbar=0.01/\sqrt{2}, \bt_i=2~(i=1,2,3)$ and $\tau_2=1$.
The blue dots represent the real part of the exact three-point function given by
eq.~\eqref{eq:Airy-3pt}, while the red solid curve is the
 late time $\tau$-scaling limit given by eq.~\eqref{eq:Airy-3pt-late}.}
\label{fig:airy3pt-sff}
\end{figure}

This late time approximation \eqref{eq:Airy-3pt-late}
is also written as
\begin{equation}
\begin{aligned}
 \left\bra\prod_{i=1}^3Z(\bt_i+\ri\tau_i\hbar^{-1})\right\ket_{\text{conn}}=F(\tau_1)
+F(\tau_2)-F(\tau_1+\tau_2),
\end{aligned} 
\label{eq:Airy-3pt-ftau}
\end{equation}
where $F(\tau)$ is given by
\begin{equation}
\begin{aligned}
 F(\tau)=\frac{1}{2\pi\hbar}\int_0^{\tau}d\tau'\frac{e^{-\frac{\bt_{\text{tot}}}{4}\tau'^2}}{\bt_{\text{tot}}}.
\end{aligned} 
\end{equation}
As we will see in \S\ref{sec:n-point}, eq.~\eqref{eq:Airy-3pt-ftau} is 
a special case of the general result
of $n$-point function in the $\tau$-scaling limit.
In Figure~\ref{fig:airy3pt-sff},
we show the plot of the three-point function as a function
of $\tau_1$ with fixed $\tau_2$.
One can see that eq.~\eqref{eq:Airy-3pt-late} is a good approximation
of the exact three-point function eq.~\eqref{eq:Airy-3pt} at late times.

Next, let us consider the case where $\sum_i \tau_i\ne0$.
In this case, it is difficult to find the analytic form
of the late time behavior of \eqref{eq:Airy-3pt}.
Instead, we can study the behavior of \eqref{eq:Airy-3pt} numerically.
As an example, let us consider the case where all $\bt_i$'s are equal
\begin{equation}
\begin{aligned}
 \left\bra Z(\bt+\ri \tau\hbar^{-1})^3\right\ket_{\text{conn}}.
\end{aligned} 
\label{eq:ary3pt-ne}
\end{equation}

\begin{figure}[t] 
\centering
\includegraphics
[width=0.9\linewidth]{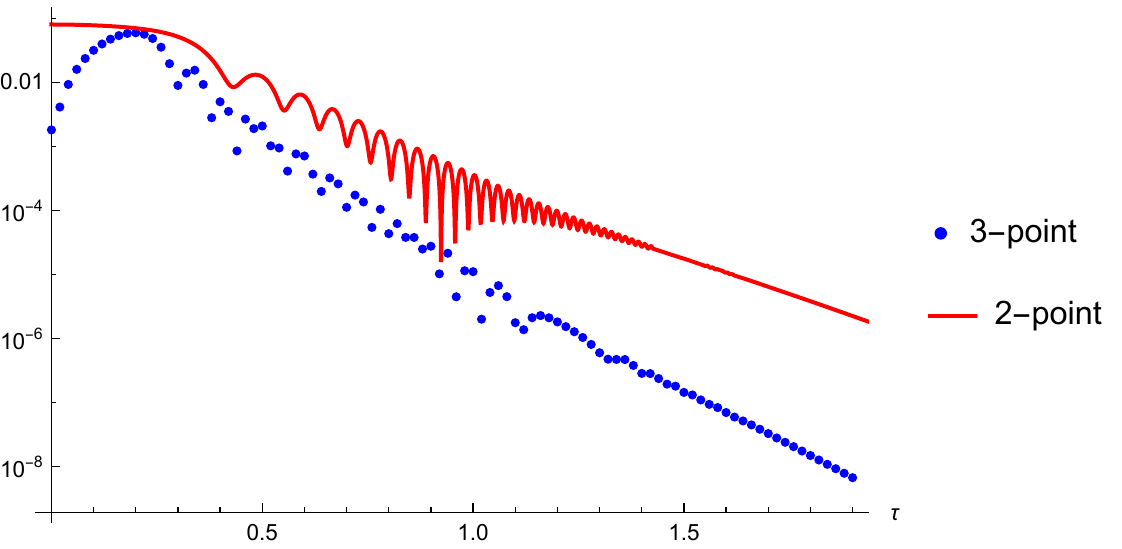}
\caption{Log-plot of the absolute value of the 
three-point function (blue) in eq.~\eqref{eq:ary3pt-ne}  
as a function of $\tau$, where $\sum_i\tau_i \neq 0$. 
For comparison, we also log-plot the absolute value of the 
two-point function (red) in eq.~\eqref{twopointconnbeta12} with $\beta_1
 = \beta_2 = \beta + \ri \tau \hbar^{-1}$.  Both show 
exponential decay. 
In this figure we set $\hbar=0.01/\sqrt{2}$ and $\bt=2$.}
\label{fig:airy3pt-dacey}
\end{figure}

As we can see from Figure~\ref{fig:airy3pt-dacey}, 
both three-point function in eq.~\eqref{eq:ary3pt-ne} and two point function eq.~\eqref{twopointconnbeta12} with $\beta_1 = \beta_2 = \beta + \ri \tau \hbar^{-1}$ 
decay exponentially as a function of $\tau$. This is in contrast to the one-point function which decays $\sim e^{ - \tau^2}$ rather than $\sim e^{- \tau}$.

\section{One point SFF  
in general topological gravity}\label{sec:1pt}

\subsection{Preliminaries}

In this section we consider general WK topological gravity one point function. Especially we consider 
the absolute square of the one point function, which is equivalent to a {disconnected} two-point function. 
When the general couplings $t_k$ are turned on,
the correlation functions can no longer be expressed
in a closed form. In certain specific regimes of the parameters,
however, one can still study them analytically
by series expansion.

We briefly recall how the one point function is expanded
\cite{Okuyama:2019xbv,Okuyama:2020ncd}.
The correlation functions for general topological gravity
are neatly expressed in terms of
the Itzykson--Zuber variables \cite{Itzykson:1992ya}
\begin{align}
\label{eq:defIn}
I_n=
I_n(u_0,\{t_k\})=\sum_{\ell=0}^\infty t_{n+\ell}\frac{u_0^\ell}{\ell!}
\quad (n\ge 0),\qquad s:=1-I_1.
\end{align}
Here $-u_0=E_0$ is the threshold energy.
It is zero for the Airy and JT gravity cases,
but for the general case
it is given by a formal power series
in $t_k$ determined by the genus zero string equation
\begin{align}
u_0=I_0(u_0,\{t_k\}).
\end{align}

In what follows let us present three different regimes
where exact expansion is available.
First, let us consider the regime
\begin{align}
\hbar\to 0,\qquad \beta:\mbox{finite.}
\end{align}
In this regime the one-point function can be expanded as
\cite{Okuyama:2019xbv,Okuyama:2020ncd}
\begin{align}
\hspace{-4mm} \bra Z(\beta) \ket
 =\sqrt{\frac{\beta}{2\pi}}e^{\beta u_0}
 \sum_{g=0}^\infty (\sqrt{2}\hbar)^{2g-1}Z_{g,1},
\label{eq:Z1exp}
\end{align}
where
\begin{align}
\begin{aligned}
Z_{0,1}
 =e^{-\beta u_0}
  \int_{-\infty}^{u_0}dv\left(I_0(v,\{t_k\})-v\right)e^{\beta v} \,,
\quad
Z_{1,1}
 =\frac{I_2}{24s^2}+\frac{\beta}{24s} \,,\quad  \cdots.
\end{aligned}
\end{align}

Next, let us consider what we call the low-temperature limit
\begin{align}
\hbar\to 0,\quad\beta\to\infty\quad\mbox{with}\quad
h:=\hbar \, \beta^{3/2}\quad \mbox{fixed}.
\end{align}
In this regime $\bra Z(\beta)\ket$ can be expanded as
\cite{Okuyama:2019xbv}
\begin{align}
\bra Z(\beta) \ket
 =\frac{e^{\frac{h^2}{12s^2}+\frac{u_0}{T}}}{2\sqrt{\pi}h}
  \sum_{\ell=0}^\infty\frac{T^\ell}{\ell!}z_\ell,
\label{eq:Z1low}
\end{align}
where $T=\beta^{-1}$ and
\begin{align}
z_0=s,\qquad
z_1=\left(1+\frac{h^4}{60s^4}\right)I_2,\qquad
\cdots.
\end{align}

Third, we can also consider what we call the 't Hooft limit
\begin{align}
 \hbar\to 0,\quad\beta\to\infty\quad
\mbox{with}\quad\lambda:=\hbar\beta\quad\mbox{fixed}.
\end{align}
In this regime the one point function can be expanded as
\cite{Okuyama:2021cub}
\begin{align}
\begin{aligned}
\bra Z(\beta) \ket
 &=\exp\left[\sum_{n=0}^\infty\hbar^{n-1}\cF_n(\lambda)\right],
\end{aligned}
\label{eq:Z1tHooft}
\end{align}
where
\begin{align}
\begin{aligned}
\cF_0
&=u_0\lambda
 +\sum_{\substack{j_a\ge 0\\[.5ex]
                   \sum_a j_a=k\\[.5ex]
                   \sum_a aj_a=n}}
 \frac{(2n+k+1)!}{(2n+3)!}\frac{\lambda^{2n+3}}{2^{n+1}s^{2n+k+2}}
 \prod_{a=1}^\infty\frac{I_{a+1}^{j_a}}{j_a!(2a+1)!!^{j_a}}\\
&=u_0\lambda
 +\frac{1}{12s^2}\lambda^3
 +\frac{I_2}{60s^5}\lambda^5
 +\left(\frac{I_2^2}{144s^8}+\frac{I_3}{840s^7}\right)\lambda^7
 +{\cal O}(\lambda^9),\\
\cF_1
 &=\frac{1}{2}\log\left[\frac{\hbar}{32\pi}
     \frac{\partial_\lambda\xi_*}{(\xi_*-u_0)^2}\right],
\qquad \xi_*(\lambda)=\partial_\lambda\cF_0,\qquad \cdots.
\end{aligned}
\end{align}

By using these results, we will evaluate
the one point function at two different time scales
in the following subsections.

\subsection{One point function at $t\sim\hbar^{-2/3}$}

Let us first evaluate the one-point function
in the regime $t\sim \hbar^{-2/3}$.
This corresponds to the low-temperature limit
and we can use the expansion \eqref{eq:Z1low}.
As in the Airy case, we consider the absolute square 
 corresponding to the disconnected two-point function
\begin{align}
\left| \bra Z(\beta+\ri t) \ket \right|^2
=\bra Z(\beta+\ri t) \ket  \bra Z(\beta-\ri t) \ket.
\label{eq:Z1squared}
\end{align}
Given the low-temperature expression eq.~\eqref{eq:Z1low}
we replace the variables as
\begin{align}
\beta\to\beta+\ri t=\beta+\frac{\ri \tau}{\hbar^{2/3}}.
\label{h2/3tauscaling}
\end{align}
Here $\beta$ and $\tau$ are supposed to be of the order of $\hbar^0$.
We see that
\begin{align}
\begin{aligned}
h^2&=\hbar^2\left(\beta+\ri \tau\hbar^{-2/3}\right)^3\\
 &=-\ri \tau^3-3\beta\tau^2\hbar^{2/3}+3\beta^2 \ri \tau\hbar^{4/3}
 +\beta^3\hbar^2,\\
T&=(\beta+\ri \tau\hbar^{-2/3})^{-1}
 =\frac{\hbar^{2/3}}{\ri \tau}+{\cal O}(\hbar^{4/3}).
\end{aligned}
\end{align}
Substituting these into \eqref{eq:Z1low} we have
\begin{align}
\left\bra  Z\left(\beta+\frac{\ri \tau}{\hbar^{2/3}}\right) \right\ket 
\left\bra  Z\left(\beta-\frac{\ri \tau}{\hbar^{2/3}}\right) \right\ket 
=\frac{e^{2\beta u_0 
 -\beta\tau^2\hbar^{2/3}/2s^2+\beta^3\hbar^2/6s^2} } 
      {4\pi[\tau^3+{\cal O}(\hbar^{2/3})]}
 \left[s^2+{\cal O}(\hbar^{2/3})\right].
\label{eq:Z1LT}
\end{align}
This expression is valid as long as $\tau\sim \hbar^0$.
We see that at the leading order of $\hbar$ expansion,
the one point function shows a power law scaling decay in $\tau$ as 
\begin{align}
\left\bra Z\left(\beta+\frac{\ri \tau}{\hbar^{2/3}}\right) \right\ket
\left\bra Z\left(\beta-\frac{\ri \tau}{\hbar^{2/3}}\right) \right\ket
=\frac{s^2 e^{2\beta u_0}}
      {4\pi \tau^3}
+{\cal O}(\hbar^{2/3})  \,.
\label{eq:Z1LT2}
\end{align}

Naively, one can see from eq.~\eqref{eq:Z1LT} that
the exponential decay, as a function of $\tau$,
becomes relevant
when $\tau\sim\hbar^{-1/3}\beta^{-1/2}$.
In terms of the original time $t=\tau/\hbar^{2/3}$,
this means $t\sim\hbar^{-1}\beta^{-1/2} \sim t_{\rm higher \, genus}$.
This is seemingly
in accordance with the analysis in the Airy case on eq.~\eqref{defthiger}.
However, careful analysis is required
because for $\tau\sim\hbar^{-1/3}$ the parameter $h$ actually
diverges as $h^2 \sim\hbar^{-1}$ and the expansion in eq.~\eqref{eq:Z1low}
no longer makes sense.
We will therefore study this time scale
using a different expansion in the next subsection.

\subsubsection{Airy and JT gravity one point function at $t \sim \hbar^{-2/3}$}
Let us look into some concrete examples.
\begin{enumerate}
\item In the Airy case, where $t_k=0$ for all $k$, we have
\begin{align}
u_0=0,\quad s=1,\quad I_{k\ge 2}=0.
\end{align}
\item In the JT gravity case, we have
\begin{align}
t_0=t_1=0,\quad
t_k=\frac{(-1)^k}{(k-1)!}\quad(k\ge 2),
\end{align}
which means
\begin{align}
u_0=0,\quad
s=1,\quad
I_k=\frac{(-1)^k}{(k-1)!}\quad(k\ge 2).
\end{align}
\end{enumerate}
Therefore, in both cases eq.~\eqref{eq:Z1LT2} becomes
\begin{align}
\left\bra Z\left(\beta+\frac{\ri \tau}{\hbar^{2/3}}\right) \right\ket
\left\bra Z\left(\beta-\frac{\ri \tau}{\hbar^{2/3}}\right) \right\ket
=\frac{1}{4 \pi \tau^3} + {\cal O}(\hbar^{2/3}).
\end{align}
Of course, this is consistent with the Airy gravity case, where we have eq.~\eqref{planaranalyticcontin} with the low-temperature scaling given by eq.~\eqref{h2/3tauscaling}. 
The difference between Airy gravity and JT gravity starts appearing only
at the higher orders 
in $\hbar$.

\subsection{One point function at $t\sim\hbar^{-1}$}

Let us next evaluate \eqref{eq:Z1squared}
at the time scale $t\sim\hbar^{-1}$,
which is known as the $\tau$-scaling limit
\cite{Saad:2022kfe,Blommaert:2022lbh,Weber:2022sov}.
This is done by substituting
\begin{align}
\lambda= \beta\hbar \pm \ri \tau
\end{align}
into the 't Hooft expansion eq.~\eqref{eq:Z1tHooft}
and then re-expanding it in $\hbar$.
We obtain
\begin{align}
\left\bra Z\left(\beta+\frac{\ri \tau}{\hbar}\right) \right\ket 
\left\bra Z\left(\beta-\frac{\ri \tau}{\hbar}\right) \right\ket
&=\frac{\hbar}{32\pi}
  \frac{\partial_\tau E_\tau}{(E_\tau-E_0)^2}
  e^{-2\beta E_\tau}+{\cal O}(\hbar^2),
\label{eq:ZZtHooft}
\end{align}
where
\begin{align}
\begin{aligned}
E_\tau&=-\xi_*(\ri\tau)\\
&=-u_0
 +\sum_{\substack{j_a\ge 0\\[.5ex]
                   \sum_a j_a=k\\[.5ex]
                   \sum_a aj_a=n}}
 \frac{(2n+k+1)!}{(2n+2)!}\frac{(-1)^n\tau^{2n+2}}{2^{n+1}s^{2n+k+2}}
 \prod_{a=1}^\infty\frac{I_{a+1}^{j_a}}{j_a!(2a+1)!!^{j_a}}\\
 &=E_0
 +\frac{1}{4s^2}\tau^2
 -\frac{I_2}{12s^5}\tau^4
 +\left(\frac{7I_2^2}{144s^8}+\frac{I_3}{120s^7}\right)\tau^6
 +{\cal O}(\tau^8).
\end{aligned}
\label{eq:Etau}
\end{align}
Therefore we see that the one point function shows an {exponential decay} at $t\sim \hbar^{-1}$, 
controlled by the function $E_\tau$. This exponential decay is in contrast with the power law decay we have seen in eq.~\eqref{eq:Z1LT2} at $t \sim \hbar^{-2/3}$.  
This $E_\tau$ ubiquitously appears
in the study of the spectral form factor in the $\tau$-scaling limit
\cite{Okuyama:2023pio}.

\subsubsection{Late time behavior in Airy and  
JT gravity}
\begin{figure}[tb]
\centering
\includegraphics
[width=0.8\linewidth]{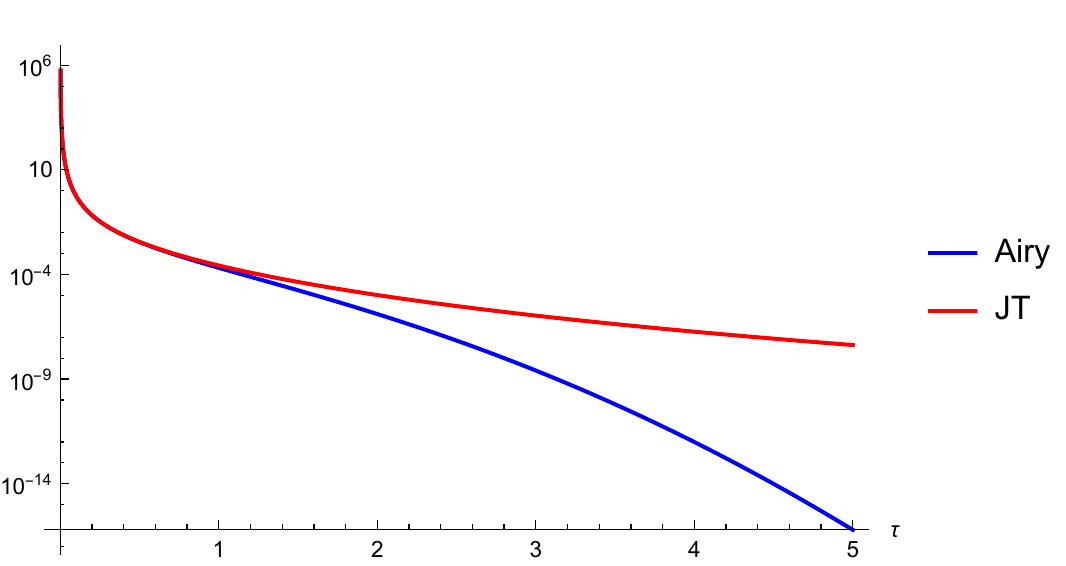}
\caption{Log-plot of 
the one-point function in the $\tau$-scaling limit for Airy gravity eq.~\eqref{eq:ZZtHooftAiry}
and JT gravity eq.~\eqref{JTdisc}. Compared with Airy gravity,
JT gravity one point function decays much more mildly. 
We set $\hbar=0.01/\sqrt{2}$ and $\bt=2$.}
\label{fig:tauscalingAiryJT}
\end{figure}

Here we look into concrete examples.
\begin{enumerate}
\item In the Airy case, we have
\begin{align}
\cF_0(\lambda)=\frac{\lambda^3}{12},\qquad
E_\tau=\frac{\tau^2}{4}, \qquad E_0=0 
\label{eq:EtauAiry}
\end{align}
and eq.~\eqref{eq:ZZtHooft} becomes
\begin{align}
\left\bra Z\left(\beta+\frac{\ri \tau}{\hbar}\right) \right\ket 
\left\bra Z\left(\beta-\frac{\ri \tau}{\hbar}\right) \right\ket
=\frac{\hbar}{4\pi\tau^3}e^{-\frac{1}{2}\beta\tau^2}+{\cal O}(\hbar^2).
\label{eq:ZZtHooftAiry}
\end{align}
This is exactly what we can obtain from the exact formula eq.~\eqref{fullanalyticconin} with $\tau$-scaling limit $t = {\tau}/{\hbar} \gg \beta$. 
\item In the JT gravity case,
we have
\begin{align}
\begin{aligned}
\cF_0(\lambda)&=\frac{1}{4}\lambda\arcsin(\lambda)^2
 +\frac{1}{2}\left(\sqrt{1-\lambda^2}\arcsin\lambda-\lambda\right),\\
E_\tau &=\frac{1}{4}\arcsinh(\tau)^2 \,,\\
E_0& =0 
\end{aligned}
\label{eq:EtauJT}
\end{align}
and eq.~\eqref{eq:ZZtHooft} becomes
\begin{align}
\left\bra Z\left(\beta+\frac{\ri \tau}{\hbar}\right) \right\ket 
\left\bra Z\left(\beta-\frac{\ri \tau}{\hbar}\right) \right\ket
&=\frac{\hbar}{4\pi}\frac{1}{\sqrt{1+\tau^2}\arcsinh(\tau)^3}
 e^{-\frac{1}{2}\beta\arcsinh(\tau)^2}\nonumber\\
&\hspace{1em}
+{\cal O}(\hbar^2).
 \label{JTdisc}
\end{align}
\end{enumerate}
For small $\tau$, 
 there is not much difference between Airy gravity and JT gravity  since
 $\arcsinh(\tau) = \tau + O(\tau^3)$. However at late time
$\tau \gtrsim 1$, the difference appears more significantly, see Figure
 \ref{fig:tauscalingAiryJT}. In the Airy case,
the one point function decays
 exponentially. However in the JT gravity case, since
 $\arcsinh (\tau)$ does not grow much and  $\arcsinh(\tau) \sim \log
 \tau$ at large $\tau$, their exponential decay becomes much
milder compared with the Airy case.    

\section{\mathversion{bold}$n$-point
SFF in general topological gravity at $t\sim\hbar^{-1}$}
\label{sec:n-point}

\subsection{Integral expression of $n$-point function}

In this section we consider the $n$-point function
at the time scale $t\sim\hbar^{-1}$.
As we saw in the last section, we can make use of the results of
the 't Hooft expansion \cite{Okuyama:2020ncd}
to study the correlation functions at this time scale.
To do this, let us first
recall that the $n$-point function can be expressed as
\cite{Okuyama:2018aij}
\begin{align}
\begin{aligned}
Z_n(\{\beta_i\})
 &:=\left\langle Z(\beta_1)\cdots Z(\beta_n)\right\rangle_{\rm conn}\\
 &=\Tr\log\left(
   1+\left[-1+\prod_{i=1}^n(1+\xi_i e^{\beta_iQ})\right]\Pi\right)
   \Bigg|_{{\cal O}(\xi_1\cdots \xi_n)}\\
 &=\Tr\log\left(
   1+\sum_{k=1}^n\sum_{i_1<\cdots<i_k}\xi_{i_1}\cdots \xi_{i_k}
           e^{(\beta_{i_1}+\cdots\beta_{i_k})Q}\Pi\right)
   \Bigg|_{{\cal O}(\xi_1\cdots \xi_n)}.
\end{aligned}
\label{eq:Zngen}
\end{align}
Here\footnote{Here we use the same normalization of $x$
as in \cite{Okuyama:2020ncd}, which differs from
that in section~\ref{sec:airy} by a factor of $\hbar$.} 
\begin{align}
Q=\partial_x^2+u,\qquad
\Pi=\int_{-\infty}^x dx'|x'\rangle\langle x'|,\qquad
x=\hbar^{-1} t_0
\end{align}
and $u=u(t_k;\hbar)$ is the KdV potential
specified by the initial condition $u(t_k;0)=u_0(t_k)$
(see \cite{Okuyama:2020ncd} for the details).
The symbol $|_{{\cal O}(\xi_1\cdots \xi_n)}$
in \eqref{eq:Zngen} means that we extract the terms
linear in all $\xi_i$. 
More explicitly, 
\eqref{eq:Zngen} is written for $n=1,2,3$ as
\begin{align}
\begin{aligned}
Z_1(\beta)
 &=\Tr\left[e^{\beta Q}\Pi\right],\\
Z_2(\beta_1,\beta_2)
 &=\Tr\left[e^{(\beta_1+\beta_2)Q}\Pi
   -e^{\beta_1 Q}\Pi e^{\beta_2 Q}\Pi\right],\\
Z_3(\beta_1,\beta_2,\beta_3)
 &=\Tr\left[e^{(\beta_1+\beta_2+\beta_3)Q}\Pi
   +e^{\beta_1 Q}\Pi e^{\beta_2 Q}\Pi e^{\beta_3 Q}\Pi
   +e^{\beta_1 Q}\Pi e^{\beta_3 Q}\Pi e^{\beta_2 Q}\Pi\right.\\
 &\hspace{2.7em}\left.
   -e^{\beta_1 Q}\Pi e^{(\beta_2+\beta_3) Q}\Pi
   -e^{\beta_2 Q}\Pi e^{(\beta_3+\beta_1) Q}\Pi
   -e^{\beta_3 Q}\Pi e^{(\beta_1+\beta_2) Q}\Pi\right].
\end{aligned}
\label{eq:Z123}
\end{align}

From the above expression
we see that $Z_n$ is a sum of the trace
\begin{align}
\Tr(e^{\beta_1 Q}\Pi \cdots e^{\beta_m Q}\Pi).
\end{align}
Note that $\beta_i$ here is in general not identical to
the original $\beta_i$ appearing in \eqref{eq:Zngen}
but could be a sum of them.
$(-Q)$ is viewed as a Hamiltonian and let $|E\rangle$
denote the energy eigenstate with eigenvalue $E$.
By inserting $1=\int dE_i |E_i\rangle\langle E_i|$
in front of each $\Pi$,
the above trace is rewritten as
\begin{align}
\Tr(e^{\beta_1 Q}\Pi \cdots e^{\beta_m Q}\Pi)
=\int dE_1\cdots\int dE_m e^{-\sum_{i=1}^m\beta_i E_i}
 K_{12}K_{23}\cdots K_{m1}.
\label{eq:TreQPi}
\end{align}
Here
\begin{align}
\begin{aligned}
K_{ij}=K(E_i,E_j)
 &=\langle E_i|\Pi|E_j\rangle\\
 &=\int_{-\infty}^x dx'\langle E_i|x'\rangle\langle x'|E_j\rangle
  =\int_{-\infty}^x\psi(E_i)\psi(E_j)
\end{aligned}
\end{align}
is the Christoffel-Darboux (CD) kernel.
At the leading order of
the small $\hbar$ expansion,
it is approximated by the sine kernel\footnote{Higher order
corrections have also been studied recently \cite{Okuyama:2023pio}.}
\begin{align}
K(E_i,E_j)
 =\frac{\sin\left[\frac{1}{2\hbar}\rho_0(E)(E_i-E_j)\right]}
       {\pi(E_i-E_j)},\qquad
E=\frac{E_i+E_j}{2}.
\label{eq:sineker}
\end{align}
Here $\rho_0(E)$ is the genus zero part of the eigenvalue density.
For general topological gravity, it is given by 
\begin{align}
\rho_0(E)=\sum_{k=1}^\infty
 \frac{(-1)^k(I_k-\delta_{k,1})
 \Gamma(\frac{1}{2})}{\Gamma(k+\frac{1}{2})}
 (E-E_0)^{k-\frac{1}{2}}.
\end{align}
Note that $E_0=-u_0$.

\subsection{General prescription for $n\ge 2$}

We are interested in the $n$-point function
with $\beta_i$ replaced by $\beta_i+\ri\tau_i\hbar^{-1}$.
For generic values of $\tau_i$,
each component of the form \eqref{eq:TreQPi}
in the $n$-point function
\eqref{eq:Zngen} scales differently
and it is not straightforward to discuss the scaling behavior
in a uniform way. 
As we will see, however, under the constraint
\begin{align}
\sum_{i=1}^n\tau_i=0
\label{eq:tausum}
\end{align}
the $n$-point function shows a universal leading order behavior.
This can be thought of as
a natural generalization of the ordinary two-point SFF
to the $n$-point one and 
in what follows we assume that this constraint is imposed.

As we have seen above, the calculation of the $n$-point SFF 
boils down to evaluating the integral of the form
\begin{align}
\begin{aligned}
I[1,2,3, \ldots , m]:=&\Tr\left[e^{(\beta_1+\ri\tau_1\hbar^{-1})Q}\Pi\cdots
         e^{(\beta_m+\ri\tau_m\hbar^{-1})Q}\Pi\right]   \quad \\ 
=&\int\left(\prod_{i=1}^m
  dE_i e^{-(\beta_i+\ri\tau_i\hbar^{-1}) E_i}\right)
  K(E_1,E_2)K(E_2,E_3)\cdots K(E_m,E_1),
\end{aligned}
\label{eq:masterint}
\end{align}
where indices in the argument of $I$
represent indices for $\beta_i + \ri \tau_i \hbar^{-1}$.
$\beta_i$ and $\tau_i$ here are not always
identical to the original ones but could be a sum of them.
Correspondingly, we will use the abbreviation where
the sum of $\beta_i + \ri \tau_i \hbar^{-1}$
is represented by the sum of indices in the argument of $I$. 
For example,  
\begin{align}
I[1+2, 3] = \Tr\left[e^{(\beta_1 + \beta_2 +\ri\tau_1\hbar^{-1}+ \ri\tau_2\hbar^{-1}      )Q}
\Pi e^{(\beta_3+\ri\tau_3\hbar^{-1})Q}   \Pi\right]   .
\end{align}
With this notation two- and three-point SFFs, obtained
from eq.~\eqref{eq:Z123} by
analytic continuation $\beta_i \to \beta_i + \ri\tau_i \hbar^{-1}$,
are concisely expressed as 
\begin{align}
\begin{aligned}
Z_2
&=I[1+2]-I[1,2],\\
Z_3
&=I[1+2+3]+I[1,2,3]+I[1,3,2]
 -I[1,2+3]-I[2,3+1]-I[3,1+2].
\end{aligned}
\end{align}
Clearly, from the cyclic property of the trace we have 
\begin{align}
I[1,2,3, \ldots , m]  = I[2,3, \ldots , m, 1] = I[3, \ldots , m, 1, 2] = \cdots.
\end{align}

We stress that $\tau_i$ in eq.~\eqref{eq:masterint} obeys the constraint
\begin{align}
\sum_{i=1}^m\tau_i=0,
\label{eq:tausumbis}
\end{align}
which is inherited from eq.~\eqref{eq:tausum}.
To evaluate the integral, we introduce the variables
\begin{align}
\hbar\omega_{ij}:=E_i-E_j
\end{align}
and rewrite the exponent as
\begin{align}
\begin{aligned}
\sum_{i=1}^m(\beta_i+\ri\tau_i\hbar^{-1}) E_i
&=\sum_{i=1}^m(\beta_i+\ri\tau_i\hbar^{-1})(\hbar\omega_{im}+E_m)\\
&=\hbar^{-1}\ri\left(\sum_{i=1}^m\tau_i\right)E_m
 +\left(\sum_{i=1}^m\beta_i\right)E_m
 +\ri\sum_{i=1}^m\tau_i\omega_{im}
 +\hbar\sum_{i=1}^m\beta_i\omega_{im}\\
&=\btot E_m+\ri\sum_{i=1}^{m-1}\tau^{(i)}\omega_{i,i+1}
  +{\cal O}(\hbar).
\end{aligned}
\label{eq:Iexponent}
\end{align}
In the last equality we have used the constraint \eqref{eq:tausumbis}
and introduced the notations
\begin{align}
\begin{aligned}
\btot&:=\sum_{i=1}^m\beta_i,\\
\tau^{(i)}&:=\sum_{k=1}^i\tau_k\qquad (i=1,\ldots,m-1).
\end{aligned}
\end{align}
Using \eqref{eq:Iexponent},
let us evaluate the integral \eqref{eq:masterint}
for small $\hbar$ in the leading order approximation.
At this level, we can replace the CD kernel by
the sine kernel \eqref{eq:sineker}
and thus the integral is approximated as
\begin{align}
\begin{aligned}
I&\simeq
 \frac{1}{\hbar}\int
 dE_m\prod_{i=1}^{m-1}d\omega_{i,i+1}
 e^{-\btot E_m-\ri\sum_{i=1}^{m-1}\tau^{(i)}\omega_{i,i+1}}
\\
&\hspace{6em}\times
 \left[\prod_{i=1}^{m-1}
  \frac{\sin\left[\frac{1}{2}\rho_0\left(\frac{E_i+E_{i+1}}{2}\right)
        \omega_{i,i+1}\right]}
       {\pi\omega_{i,i+1}}
 \right]
  \frac{\sin\left[\frac{1}{2}\rho_0\left(\frac{E_1+E_m}{2}\right)
        \sum_{i=1}^{m-1}\omega_{i,i+1}\right]}
       {\pi\sum_{i=1}^{m-1}\omega_{i,i+1}}\\
 &\simeq
 \frac{1}{\pi^m\hbar}\int
 dE\prod_{i=1}^{m-1}d\omega_{i,i+1}
 e^{-\btot E-\ri\sum_{i=1}^{m-1}\tau^{(i)}\omega_{i,i+1}}\\
&\hspace{6em}\times
 \left[\prod_{i=1}^{m-1}
  \frac{\sin\left[\frac{1}{2}\rho_0(E)
        \omega_{i,i+1}\right]}
       {\omega_{i,i+1}}
 \right]
  \frac{\sin\left[\frac{1}{2}\rho_0(E)
        \sum_{i=1}^{m-1}\omega_{i,i+1}\right]}
       {\sum_{i=1}^{m-1}\omega_{i,i+1}}.
\end{aligned}
\label{eq:Iapprox}
\end{align}

The last integral is neatly evaluated by the method
of Lagrange multipliers.\footnote{We
would like to thank the anonymous referee for suggesting
this prescription.}
By inserting
\begin{align}
1
 =\frac{1}{2\pi}\int_{-\infty}^\infty d\omega
  \int_{-\infty}^\infty d\tau e^{-\ri\tau
  \left(\omega-\sum_{i=1}^{m-1}\omega_{i,i+1}\right)},
\end{align}
the integral \eqref{eq:Iapprox} is expressed as
\begin{align}
\begin{aligned}
I
&=\frac{1}{2\pi^{m+1}\hbar}\int
 dE d\tau d\omega\prod_{i=1}^{m-1}d\omega_{i,i+1}
 e^{-\btot E-\ri\tau\omega
    -\ri\sum_{i=1}^{m-1}(\tau^{(i)}-\tau)\omega_{i,i+1}}\\
&\hspace{6em}\times
 \left[\prod_{i=1}^{m-1}
  \frac{\sin\left[\frac{1}{2}\rho_0(E)
        \omega_{i,i+1}\right]}
       {\omega_{i,i+1}}
 \right]
  \frac{\sin\left[\frac{1}{2}\rho_0(E)
        \omega\right]}
       {\omega}.
\end{aligned}
\end{align}
As in the case of the ordinary SFF \cite{brezin-hikami,Okuyama:2023pio},
we can evaluate this sort of integral using the formula
\begin{align}
\int_{-\infty}^\infty d\omega e^{-\ri\omega\tau}
\frac{\sin a\omega}{\pi\omega}
=\theta\left(a-|\tau|\right)\qquad (a>0).
\label{eq:thetaformula}
\end{align}
Here, $\theta(x)$ is the step function.
The result is
\begin{align}
\begin{aligned}
I
&=\frac{1}{2\pi\hbar}\int dEd\tau e^{-\btot E}
  \prod_{i=0}^{m-1}
  \theta\left(\tfrac{1}{2}\rho_0(E)-|\tau-\tau^{(i)}|\right),
\end{aligned}
\end{align}
where we have formally introduced $\tau^{(0)}:=0$
for the sake of brevity.
Observe that the arguments of the step functions satisfy
\begin{align}
\left\{\begin{array}{lll}
\tfrac{1}{2}\rho_0(E)-|\tau-\tau^{(i)}|\ge
\tfrac{1}{2}\rho_0(E)-|\tau-\tau^{(\mathrm{min})}|
&\ \mbox{when}\ &
\tau\ge\tau^{(\mathrm{mid})},
\\[1ex]
\tfrac{1}{2}\rho_0(E)-|\tau-\tau^{(i)}|\ge
\tfrac{1}{2}\rho_0(E)-|\tau-\tau^{(\mathrm{max})}|
&\ \mbox{when}\ &
\tau\le\tau^{(\mathrm{mid})}
       \end{array}\right.
\end{align}
for all $i=0,\ldots,m-1$, where
\begin{align}
\begin{aligned}
\tau^{(\mathrm{min})}
&:=\min\bigl\{\tau^{(0)}=0,\tau^{(1)},\ldots,\tau^{(m-1)}\bigr\},\\
\tau^{(\mathrm{max})}
&:=\max\bigl\{\tau^{(0)}=0,\tau^{(1)},\ldots,\tau^{(m-1)}\bigr\},\\
\tau^{(\mathrm{mid})}
&:=\tfrac{1}{2}\left(\tau^{(\mathrm{min})}+\tau^{(\mathrm{max})}\right).
\end{aligned}
\end{align}
%
By using these inequalities, the integral is simplified as
\begin{align}
\begin{aligned}
I&=\frac{1}{2\pi\hbar}\int dE e^{-\btot E}
   \int_{\tau^{(\mathrm{mid})}}^\infty d\tau\,
   \theta\left(\tfrac{1}{2}\rho_0(E)-|\tau-\tau^{(\mathrm{min})}|\right)
\\
&\hspace{1em}
  +\frac{1}{2\pi\hbar}\int dE e^{-\btot E}
   \int_{-\infty}^{\tau^{(\mathrm{mid})}} d\tau\,
   \theta\left(\tfrac{1}{2}\rho_0(E)-|\tau-\tau^{(\mathrm{max})}|\right).
\end{aligned}
\end{align}
By changing the variable as $\tau=\tilde{\tau}/2+\tau^{(\mathrm{min})}$
in the first term
and $\tau=-\tilde{\tau}/2+\tau^{(\mathrm{max})}$
in the second term,
one finds that the first and second terms are in fact identical.
Summing them up, one obtains
\begin{align}
\begin{aligned}
I&=\frac{1}{2\pi\hbar}\int dE e^{-\btot E}
   \int_{\tau^{(\mathrm{max})}-\tau^{(\mathrm{min})}}^\infty d\tilde{\tau}\,
   \theta\left(\rho_0(E)-\tilde{\tau}\right)\\
&=\frac{1}{2\pi\hbar}
  \int_{\tau^{(\mathrm{max})}-\tau^{(\mathrm{min})}}^\infty d\tilde{\tau}
  \int_{E_{\tilde{\tau}}}^\infty dE e^{-\btot E}\\
&=\frac{1}{2\pi\hbar\btot}
  \int_{\tau^{(\mathrm{max})}-\tau^{(\mathrm{min})}}^\infty d\tilde{\tau}
  e^{-\btot E_{\tilde{\tau}}},
\end{aligned}
\end{align}
where in the second equality we have used the fact that
$E_\tau$ in eq.~\eqref{eq:Etau}
satisfies the equation \cite{Okuyama:2023pio}
\begin{align}
\rho_0(E_\tau)=\tau.
\label{Etaurelationship}
\end{align}
We introduce the function
\begin{align}
F(\tau):=\frac{1}{2\pi\hbar\btot}\int_0^\tau
 d\tilde{\tau} e^{-\btot E_{\tilde{\tau}}}
\label{eq:Fdef}
\end{align}
and write the result as
\begin{align}
I=F(\infty)-F(\tau^{(\mathrm{max})}-\tau^{(\mathrm{min})}).
\label{eq:Iresult1}
\end{align}

A few comments are in order.
First, $F(\tau)$ is an odd function:
\begin{align}
F(-\tau)=-F(\tau).
\label{eq:fodd}
\end{align}
This immediately follows from the fact that
$E_\tau$ is an even function in $\tau$,
as given in \eqref{eq:Etau}. 
Second, using the fact that $E_\tau$
satisfies the equation \eqref{Etaurelationship},
we can change the integration variable
and rewrite $F(\tau)$ as\footnote{We assume that
$\rho_0(E)$ grows monotonically for $E>E_0$.} 
\begin{align}
\begin{aligned}
F(\tau)
 &=\frac{1}{2\pi\hbar\btot}
   \int_{E_0}^{E_\tau} dE\frac{d\rho_0(E)}{dE}e^{-\btot E}\\
 &=\tau e^{-\btot E_\tau}
  +\frac{1}{2\pi\hbar}
   \int_{E_0}^{E_\tau}dE\rho_0(E) e^{-\btot E}.
\end{aligned}
\end{align}
From this expression it is clear that in the limit of $\tau\to\infty$,
$F(\tau)$ becomes
\begin{align}
\begin{aligned}
F(\infty)
 &=\frac{1}{2\pi\hbar}\int_{E_0}^\infty dE\rho_0(E)e^{-\btot E}\\
 &=\langle Z(\btot)\rangle_{g=0}.
\end{aligned}
\end{align}
This leads us to define the normalized function
\begin{align}
\begin{aligned}
f(\tau):=\frac{F(\tau)}{\langle Z(\btot)\rangle_{g=0}}.
\label{eq:ftaudef}
\end{aligned}
\end{align}
Clearly, $f(\tau)$ satisfies
\begin{align}
f(0)=0,\qquad \lim_{\tau\to\infty}f(\tau)=1,\qquad
f(-\tau)=-f(\tau).
\label{eq:fprop}
\end{align}
As we will see, 
this function $f(\tau)$ is essentially the connected two-point SFF. 
In terms of $f(\tau)$
the final result \eqref{eq:Iresult1} is expressed as
\begin{align}
I=\langle Z(\btot)\rangle_{g=0}
 \left(1-f(\tau^{(\mathrm{max})}-\tau^{(\mathrm{min})})\right).
\label{eq:Iresult2}
\end{align}

As an illustration we write down explicit formulas for Airy and JT gravity cases. 
\begin{itemize}
\item In the Airy case, we have $E_\tau=\tau^2/4$, $\rho_0(E)=2\sqrt{E}$ and
\begin{align}
\langle Z(\btot)\rangle_{g=0}=\frac{1}{2\hbar\sqrt{\pi\btot^3}} \,.
\end{align}
Then $f(\tau)$ is calculated as
\begin{align}
\label{ftauforairy}
f(\tau)=\mathrm{Erf}\left(\frac{\tau}{2}\sqrt{\btot}\right).
\end{align}
For the case of $\beta_1$ and $\beta_2$ given in
eq.~\eqref{beta1beta2rela},
we have $\btot = 2 \beta$  and 
\begin{align}
\hbar\sqrt{\beta^2+t^2}\to\tau\quad\mbox{for}\quad\hbar\to 0. 
\end{align} 
Then, $f(\tau)$ in eq.~\eqref{ftauforairy} matches with the connected two-point function given by eq.~\eqref{fullairyconn}.  

\item In the JT gravity case, we have $E_\tau=\arcsinh(\tau)^2/4$,
$\rho_0(E)=\sinh(2\sqrt{E})$ and
\begin{align}
\langle Z(\btot)\rangle_{g=0}
 =\frac{e^{1/\btot}}{2\hbar\sqrt{\pi\btot^3}} \,.
\end{align}
Then $f(\tau)$ is obtained as
\begin{align}
f(\tau)=
\frac{1}{2}
 \left[
  \Erf\left(\frac{\btot\arcsinh(\tau)+2}{2\sqrt{\btot}}\right)
 +\Erf\left(\frac{\btot\arcsinh(\tau)-2}{2\sqrt{\btot}}\right)
 \right].
\end{align}

In fact, this reproduces the known result of the two-point function \cite{Saad:2022kfe}. 
\end{itemize}

To sum up, we arrive at the conclusion that
the $n$-point function 
\begin{align}
Z_n\left(\beta_1+\frac{\ri\tau_1}{\hbar},\ldots,
  \beta_n+\frac{\ri\tau_n}{\hbar}\right)
\end{align}
with the constraint \eqref{eq:tausum} is given by,
in the leading order approximation,
a certain sum of the function \eqref{eq:Fdef}.
As an illustration
we will present the explicit form of the sum
for small $n$ in the next subsection.

\subsection{Several examples}

In this subsection we will present the explicit form
of the sum of $f(\tau)$ for two- and three-point functions.
We will also present the sum for four-point function
in a particular case.

\begin{itemize}
\item For the two point function, without loss of generality we can assume
\begin{align}
\tau_1=-\tau_2>0.
\end{align}
Since 
\begin{align}
\begin{aligned}
Z_2
&=I[1+2]-I[1,2], 
\end{aligned}
\end{align}
in the leading order approximation 
the first term gives
\begin{align}
I[1+2]=\Tr e^{(\beta_1+\beta_2)Q}\Pi
 =\langle Z(\btot)\rangle_{g=0}.
\end{align}
The second term is evaluated
by using the result in the previous subsection.
Obviously, we have $\tau^{(\mathrm{max})}=\tau^{(1)}=\tau_1$, 
$\tau^{(\mathrm{min})}=0$,  
and thus
\begin{align}
I[1,2]=\Tr\left[e^{(\beta_1 + \ri \tau_1 \hbar^{-1}) Q}\Pi e^{ (\beta_2 + \ri \tau_2 \hbar^{-1})Q}\Pi\right]
=\langle Z(\btot)\rangle_{g=0}\left(1-f(\tau_1)\right). 
\end{align}
In total, we obtain
\begin{align}
\label{ftauasconnectedtwopt}
Z_2=\langle Z(\btot)\rangle_{g=0}f(\tau_1) \,.
\end{align}
%
This manifestly shows that $f(\tau)$
is essentially the connected two-point SFF.

\item Next, let us consider the three point function.
Without loss of generality we can assume that
\begin{align}
\tau_1,\tau_2>0,\quad \tau_3<0.
\end{align}
Similarly, we find
\begin{align}
Z_3
&=I[1+2+3]+I[1,2,3]+I[1,3,2]
 -I[1,2+3]-I[2,3+1]-I[3,1+2] \nonumber\\
&=\langle Z(\btot)\rangle_{g=0}
 \left[
 -f(\tau_1+\tau_2)-f(\tau_2+\tau_1)
 +f(\tau_1)+f(\tau_2)+f(\tau_1+\tau_2)\right]\nonumber \\
 &=\langle Z(\btot)\rangle_{g=0}
 \sum_{i=1}^3f(\tau_i).
\label{eq:Z3result}
\end{align}
The third equality
we have used $f(\tau_1+\tau_2)=f(-\tau_3)=-f(\tau_3)$.
Using the second property of $f(\tau)$ in eq.~\eqref{eq:fprop},
one can see that the plateau is correctly reproduced
\begin{align}
\lim_{\tau_1,\tau_2\to\infty}Z_3=\langle Z(\btot)\rangle_{g=0}.
\end{align}
Eq.~\eqref{eq:Z3result} reproduces the three point result
presented in eq.(6.9) of \cite{Blommaert:2022lbh}. 

\item Let us next consider the four-point function.
From the general formula \eqref{eq:Zngen} one obtains
\begin{align}
Z_4
&=I[1+2+3+4]\nonumber\\
&\hspace{1em}
 -I[1,2+3+4]-I[2,3+4+1]-I[3,4+1+2]-I[4,1+2+3]\nonumber\\
&\hspace{1em}
 -I[1+2,3+4]-I[1+3,2+4]-I[1+4,2+3]\nonumber\\
&\hspace{1em}
 +2(I[1,2,3+4]+I[1,3,2+4]+I[1,4,2+3]\nonumber\\
&\hspace{3em}
 +I[2,3,1+4]+I[2,4,1+3]+I[3,4,1+2])\nonumber\\
&\hspace{1em}
 -6I[1,2,3,4].
\label{eq:Z4inI}
\end{align}
In this paper we only consider the case
\begin{align}
\tau_1,\tau_2,\tau_3>0,\quad \tau_4<0.
\end{align}
%
In this case, for every integral $I$
in \eqref{eq:Z4inI} one can take $\tau^{(i)}>0$,
so that $\tau^{(\mathrm{min})}=0$
and $\tau^{(\mathrm{max})}$ is unambiguously determined.
In the same way as above, we obtain 
\begin{align}
\begin{aligned}
Z_4
&=\langle Z(\btot)\rangle_{g=0}\\
&\hspace{1em}
\times[
  f(\tau_1)+f(\tau_2)+f(\tau_3)+f(\tau_1+\tau_2+\tau_3)\\
&\hspace{3em}
 +f(\tau_1+\tau_2)+f(\tau_1+\tau_3)+f(\tau_2+\tau_3)\\
&\hspace{3em}
 -2\bigl(f(\tau_1+\tau_2)+f(\tau_1+\tau_3)+f(\tau_1+\tau_2+\tau_3)\\
&\hspace{4em}
 +f(\tau_2+\tau_3)+f(\tau_2+\tau_1+\tau_3)+f(\tau_3+\tau_1+\tau_2)\bigr)\\
&\hspace{3em}
 +6f(\tau_1+\tau_2+\tau_3)]\\[1ex]
&=\langle Z(\btot)\rangle_{g=0}\\
&\hspace{1em}\times
 \left[\sum_{i=1}^3f(\tau_i)
 -f(\tau_1+\tau_2)-f(\tau_2+\tau_3)-f(\tau_3+\tau_1)
 +f(\tau_1+\tau_2+\tau_3)\right].
\end{aligned}
\end{align}
Using the second property of $f(\tau)$ in eq.~\eqref{eq:fprop},
one can see that the plateau is correctly reproduced
%
\begin{align}
\lim_{\tau_1,\tau_2,\tau_3\to\infty}Z_4=\langle Z(\btot)\rangle_{g=0}.
\end{align}
\end{itemize}

One can consider other cases as well
and evaluate the integral using the prescription in the last
subsection.
While there is no technical difficulty, one needs to
handle many cases separately in order to fix $\tau^{(\mathrm{max})}$
and $\tau^{(\mathrm{min})}$ 
for each constituent integral, which is rather laborious.


\section{Summary and discussion}
\label{sec:discussion} 
In this paper, we have analyzed
$n$-point SFFs in WK topological gravity,
and as its special cases, in Airy and JT gravities in detail. 
While $2$-point SFF of JT gravity
 was  studied recently \cite{Blommaert:2022lbh, Weber:2022sov, Saad:2022kfe,
Okuyama:2023pio}, our focus has been on the late time behavior of the $n$-point SFFs, 
especially on the two typical timescales, $t\sim \hbar^{-2/3}$ and $t\sim
\hbar^{-1}$.  
Moreover, for Airy gravity we have done full analysis at all
time-scale using the exact result by Okounkov \cite{Okounkov:2001usa}.

Regarding one point SFF (or equivalently disconnected two point SFF), 
we have found that it decays by power law at $t\sim\hbar^{-2/3}$ ($\hbar \to 0$).
However at much later time $t\sim \hbar^{-1}$, it decays exponentially. 
For connected two-point SFF in Airy gravity, we have found 
that it first shows the $t$-linear ramp behavior, which changes into plateau at 
$t=t_{\rm higher \, genus}\sim\hbar^{-1}$. The dip time, where the disconnected and connected two-point SFFs become 
the same order, is at $t \sim \hbar^{-1/2}$. These are the same behavior seen in JT gravity \cite{Okuyama:2020ncd, Blommaert:2022lbh, Saad:2022kfe, Okuyama:2023pio}. 
Not only one- and two-point SFFs, we have discussed general $n (\ge 2)$-point SFF at $t \sim \tau \hbar^{-1}$ with $\tau$ fixed. 
The crucial point in this late time is that  
$n (\ge 2)$-point SFF is characterized by a single function $F(\tau)$ or its normalized form $f(\tau)$ given by eq.~\eqref{eq:Fdef} or \eqref{eq:ftaudef}, respectively.  
$F(\tau)$ is characterized by $E_\tau$ given in \eqref{eq:Etau},
which is
implicitly determined by the classical eigenvalue
distribution $\rho_0(E)$ through eq.~\eqref{Etaurelationship}. 
$f(\tau)$ essentially represents the connected two-point SFF. 
Therefore, general $n \ge 2$-point SFFs at late time $t \sim \hbar^{-1}$ are characterized
 by its connected two-point SFF and are specified by the 
classical eigenvalue density. 
 These suggest that 
the qualitative behaviors of general $n (\ge 2)$-point SFFs in Airy and JT gravities are very similar. 
From the viewpoint of general topological gravity, these theories 
 correspond to different values of parameters $t_k$ yielding different classical eigenvalue densities. Therefore the qualitative behavior mentioned above seems universal, independent of the particular choices of $t_k$ for topological gravity.

The essential point is that the connected two point SFF determines
the multiple $n$-point SFFs. 
This is due to the fact that the $n$-point SFFs are determined by the Christoffel-Darboux (CD) kernel.   
Especially at the late time $t \sim \hbar^{-1}$, CD kernel is
approximated by the sine-kernel which represents the eigenvalue repulsion. 
This sine-kernel is essential for the ramp and plateau. 
The fact that general $n$-point SFFs are characterized by the sine-kernel is due to 
the equivalence of WK topological gravity with the one-matrix model \cite{Witten:1990hr,Kontsevich:1992ti}. In one-matrix
model, one can always diagonalize it
with Vandermonde determinant, which is the origin of the eigenvalue repulsion. In this way,
general $n$-point SFF inherits the characteristics of two point SFF.

These imply that topological gravity which is dual to the one-matrix model is too simple in some sense since two-point SFF is all the non-trivial information we need. 
To go beyond topological gravity and to study more realistic gravity as
is higher dimensional one, one might need to study multi-matrix
models. See for the recent approach along these lines \cite{Jafferis:2022wez}. However, these multi-matrices are difficult to solve analytically in general.  
On the gravity side,
JT gravity can acquire bulk degrees of freedom if one includes matter fields. 
However, the long thin tube limit of the moduli space show divergence
due to the negative Casimir energy \cite{Saad:2019lba,
Moitra:2021uiv, Moitra:2022glw}. 
It is very interesting to study how the correspondence between JT gravity with matters and multi-matrix
models works in such cases and 
figure out how to cure these divergences.

\acknowledgments
We would like to thank the anonymous referee
for several helpful suggestions. This work was supported in part by JSPS KAKENHI Grant Number 21J20906 (TA), 18K03619 (NI), 19K03856 (KS), 22K03594 (KO). This work was also supported by MEXT KAKENHI Grant-in-Aid for Transformative Research Areas A ``Extreme Universe'' No.~21H05184 (NI) and  No.~21H05187 (KO).

\bibliography{paper}
\bibliographystyle{utphys}

\end{document}